\documentstyle[12pt]{article}
\begin{document}
\vspace*{2cm}
\begin{center}
{\large \bf On the spectrum of S=1/2 XXX Heisenberg chain with
elliptic exchange}\\
\vspace*{1.5cm}
{\sc V.I. Inozemtsev}\\
{\it Institute for Solid State Physics, University of Tokyo,
Tokyo 106, Japan\\
and\\
Lab. of Theor. Phys., JINR, 141980 Dubna, Russia}\\
\vspace*{1cm}
Abstract
\end{center}
\vspace*{1cm}
It is found that the Hamiltonian of S=1/2 isotropic Heisenberg chain with $N$
 sites and elliptic
non-nearest-neighbor exchange is diagonalized in each sector of the Hilbert
space with magnetization $N/2-M$, $1<M\leq[N/2]$, by means of double
 quasiperiodic meromorphic
solutions to the $M$-particle quantum Calogero-Moser problem on a line.
The spectrum and highest-weight states are determined by the solutions
of the systems of transcendental equations of the Bethe-ansatz type
which arise as restrictions to particle pseudomomenta.

\newpage
In recent years, much attention has been paid to studies of 1D lattice
systems, due to their relevance to principal notions of field theory and
experimental investigations of effectively low-dimensional crystals. Even the
simplest lattice systems, namely isotropic S=1/2 Heisenberg chains,
have unveiled rich structure and provided nontrivial examples
of many-body interactions. The corresponding mathematical problem consists
in finding the proper analytic tool for the diagonalization of the model
Hamiltonian
$${\cal H}^{(s)}={J\over4}\sum_{1\leq{j}\neq{k}\leq{N}}h(j-k)
(\vec\sigma_{j}\vec\sigma_{k}-1) \qquad h(j)=h(j+N)
\eqno(1)$$
where $\vec \sigma_{j}$ are Pauli matrices acting on spin at $j$th site.

At finite $N$, it has been successfully treated in the integrable cases
of nearest-neighbor coupling solved by Bethe [1]
$$h(j)= \delta_{\vert j({\rm mod}N)\vert,1}
\eqno(2)$$
and long-range trigonometric exchange proposed independently by Haldane and
 Shastry [2]
$$h(j)=\left({N\over\pi}\sin{{\pi j}\over{N}}\right)^{-2}.
\eqno(3)$$
At present, a number of impressive results are known for both these models.
In particular, they include the additivity of the spectrum under proper
choice of "rapidity" variables [1,3], the
 description of underlying symmetry [4,5], construction of thermodynamics
in the limit $N\to\infty$ [6,3], the connection to the continuum integrable
many-body problems [7,2], and closed-form expressions of correlations
in the antiferromagnetic ground state. The rich collection
of various generalizations and physical applications of Bethe and
Haldane-Shastry models can be found in recent review papers [8,9].

Several years ago, I have introduced a more general one-parametric form
of spin exchange which provides another example of integrable lattice
Hamiltonian (1) [10]. It has been motivated by the similarity of the
Lax representation of the Heisenberg equations of motion for continuum and
lattice models. In the former case, the most general translationally-invariant
integrable Hamiltonian with elliptic pairwise particle interaction has been
found by Calogero [11] and Moser [12],
$$H_{CM}={1\over2}\left[-\sum_{\beta=1}^{L}
{{\partial^2}\over{\partial x_{\beta}^{2}}}+\lambda(\lambda+1)
\sum_{\beta\neq\gamma}
^{L}\wp(x_{\beta}-x_{\gamma})\right].
\eqno(4)$$
The existence of extra integrals of motion commuting with (4)
has been demonstrated in [13]. Recently, the eigenvalue problem for the
elliptic Calogero-Moser operator received much attention due to its relation
to the representations of double affine algebras and solutions of Knizhnik-
Zamolodchikov-Bernard equations [14,15].

The lattice analog of (4) is given by (1) with
 $$h(j)=\left({\omega\over\pi}\sin{\pi\over\omega}\right)^2
\left[\wp_{N}(j)+{2\over\omega}\zeta_{N}\left({\omega\over2}\right)\right],
\eqno(5)$$
where $\wp_{N}(x)$, $\zeta_{N}(x)$ are the Weierstrass functions defined
on the torus $T_{N}={\bf C}/{\bf Z}N+{\bf Z}\omega$, $\omega=i\kappa$,
$\kappa\in{\bf R}_{+}$. Remarkably, it turned out that the exchange (5)
comprises both (2) and (3) [10]: in fact,
the factor in (5) is chosen as to reproduce the nearest-neighbor coupling
under periodic boundary conditions (2) in the limit $\kappa\to 0$
and the long-range exchange (3) in the limit $\kappa\to\infty$.

However, till now much less is known about the lattice model with the exchange
(5) in comparison with its limiting forms due to the mathematical complexities
caused by the presence of the Weierstrass functions. The family of the
operators
which commute with ${\cal H}^{(s)}$ has been found only recently [16]. The
simpler case of infinite chain $N\to\infty$, $h(j)\to [\sinh(\pi/\kappa)/\sinh
(\pi j/\kappa)]^{2}
$ has been considered in detail in [17]. As for finite $N$, the description
of the spectrum has been performed only for simplest two- and three-magnon
excitations over ferromagnetic vacuum [10, 18, 19].

The aim of this Letter is to demonstrate the remarkable correspondence between
the highest-weight eigenstates of the lattice Hamiltonian  with the
elliptic exchange (5) and double quasiperiodic  meromorphic eigenfunctions of
the Calogero-Moser operator (4) which allows to formulate the equations of the
 Bethe-ansatz type for calculating the whole spectrum.

The Hamiltonian (1) commutes with the operator of total spin $\vec
{\bf S}={1\over2}
\sum_{j=1}^{N}\vec \sigma_{j}$. Then the eigenproblem for it is
decomposed into the problems in the subspaces formed by the common eigenvectors
of ${\bf S}_{3}$ and $\vec {\bf S}^2$ such that $S=S_{3}=
N/2-M$, $0\leq{M}\leq[N/2]$,
$${\cal H}^{(s)}\vert\psi^{(M)}>=E_{M}\vert\psi^{(M)}>.
\eqno(6)$$
The eigenvectors $\vert \psi^{(M)}>$ are written in the usual form
$$\vert\psi^{(M)}>=\sum_{n_{1}..n_{M}}^{N}\psi_{M}(n_{1}..n_{M})
\prod_{\beta=1}^{M}
s_{n_{\beta}}^{-}\vert0>,
\eqno(7)$$
where $\vert0>=\vert\uparrow\uparrow...\uparrow>$ is the ferromagnetic ground
 state with all spins up and the summation is taken over all combinations of
 integers $\{n\}\leq N$ such that
$\prod_{\mu<\nu}^{M}(n_{\mu}-n_{\nu})\neq0$. The substitution of (7) into
(6) results in the lattice Schr\"odinger equation for completely symmetric
wave function $\psi_{M}$
$$\sum_{s\neq n_{1},..n_{M}}^{N}\sum_{\beta=1}^{M}\wp_{N}(n_{\beta}-s)
\psi_{M}(n_{1},..n_{\beta-1},s,n_{\beta+1},..n_{M})$$
$$+
\left[\sum_{\beta\neq\gamma}^{M}\wp_{N}(n_{\beta}-n_{\gamma})-{\cal E}_{M}
\right]\psi_{M}(n_{1},..n_{M})=0.
\eqno(8)$$
The eigenvalues $\{E_{M}\}$ are given by
$$E_{M}=J\left({\omega\over\pi}\sin{\pi\over\omega}\right)^{2}
\left\{{\cal E}_{M}+{2\over\omega}\left[{{2M(2M-1)-N}\over4}\zeta_{N}\left(
{\omega\over2}\right)-M\zeta_{1}\left({\omega\over2}\right)\right]
\right\},
\eqno(9)$$
where $\zeta_{1}(x)$ is the Weierstrass zeta function defined on the torus
$T_{1}={\bf C}/{\bf Z}+ {\bf Z}\omega$.

To find the solutions to (8), let us consider the following ansatz for
$\psi_{M}$:
$$
\psi_{M}(n_{1},..n_{M})=\sum_{P\in\pi_{M}}\varphi_{M}^{(p)}(n_{P1},..n_{PM}),
\eqno(10)$$
$$\varphi_{M}^{(p)}(n_{1},..n_{M})=\exp\left(-i\sum_{\nu=1}^{M}p_{\nu}n_{\nu}
\right)
\chi_{M}^{(p)}(n_{1},..n_{M}),
\eqno(11)$$
where $\pi_{M}$ is the group of all permutations $\{P\}$ of the numbers from
1 to $N$
and $\chi_{M}^{(p)}$ is the solution to the {\it continuum} quantum
many-particle problem
$$\left[-{1\over2}\sum_{\beta=1}^{M}{{\partial^2}\over{\partial x_{\beta}^2}}
+\sum_{\beta\neq\lambda}^{M}\wp_{N}(x_{\beta}-x_{\lambda})-{\sf
E}_{M}(p)\right]
\chi_{M}^{(p)}(x_{1},..x_{M})=0.
\eqno(12)$$
It is specified up to a normalization factor by the particle pseudomomenta
$(p_{1},..p_{M})$. The standard argumentation of the Floquet-Bloch theory shows
that due to perodicity of the potential term in (3) $\chi_{M}^{(p)}$ obeys
the quasiperiodicity conditions [18]
$$\chi_{M}^{(p)}(x_{1},..x_{\beta}+N,..x_{M})=\exp(ip_{\beta}N)
\chi_{M}^{(p)}(x_{1},..x_{M}),
\eqno(13)$$
$$\chi_{M}^{(p)}(x_{1},..x_{\beta}+\omega,..x_{M})=\exp(q_{\beta}(p)+ip_{\beta}
\omega)
\chi_{M}^{(p)}(x_{1},..x_{M}), \quad 0\leq{\Im}m(q_{\beta})<2\pi
\eqno(14)$$
$$
1\leq\beta\leq M.
$$
The eigenvalue ${\sf E}_{M}(p)$ is some symmetric function of
$(p_{1},..p_{M})$.
The set $\{q_{\beta}(p)\}$ is also completely determined by $\{p\}$.
In this Letter I do not refer to the explicit form of these functions
which is still unknown for $M>3$.

The structure of the singularity of $\wp_{N}(x)$ at $x=0$ implies that
$\chi_{M}^{(p)}$ can be presented in the form
$$\chi_{M}^{(p)}={{F^{(p)}(x_{1},..x_{M})}\over{G(x_{1},..x_{M})}},\quad
G(x_{1},..x_{M})=\prod_{\alpha<\beta}^{M}
\sigma_{N}(x_{\alpha}-x_{\beta}),
\eqno(15)$$
where $\sigma_{N}(x)$ is the Weierstrass sigma function on the torus $T_{N}$.
The only simple zero of $\sigma_{N}(x)$ on $T_{N}$ is located at $x=0$. Thus
$[G(x_{1},..x_{M})]^{-1}$ absorbs all the singularities of $\chi_{M}^{(p)}$
on the hypersurfaces $x_{\alpha}=x_{\beta}$. The numerator $F^{(p)}$ in (15) is
analytic on $(T_{N})^M$ and obeys the equation
$$
\sum_{\alpha=1}^{M}{{\partial^2 F^{(p)}}\over{\partial x_{\alpha}^2}}
+\left[2E_{M}(p)-{M\over2}\sum_{\alpha\neq\beta}^{M}
(\wp_{N}(x_{\alpha}-x_{\beta})-\zeta_{N}^{2}(x_{\alpha}-x_{\beta}))\right]F^{(p)}
$$
$$=\sum_{\alpha\neq\beta}\zeta_{N}(x_{\alpha}-x_{\beta})
\left({{\partial F^{(p)}}\over{\partial x_{\alpha}}}-
      {{\partial F^{(p)}}\over{\partial x_{\beta}}}\right).
\eqno(16)$$
The regularity of the left-hand side of (16) as $x_{\mu}\to x_{\nu}$
implies that
$$\left({\partial\over{\partial x_{\mu}}}-{\partial\over{\partial x_{\nu}}}
\right)
F^{p}(x_{1},..x_{M})\vert_
{x_{\mu}=x_{\nu}}=0
\eqno(17)$$
for any pair $(\mu,\nu)$.

The remarkable fact is that the properties (13-15,17) of $\chi_{M}^{(p)}$ allow
one to validate
the ansatz (10-11) for the eigenfunctions of the lattice Schr\"odinger equation
(8).
 Substitution of (10) to (8) yields
$$\sum_{P\in\pi_{M}}\left\{\sum_{\beta=1}^{M}{\cal S}_{\beta}(n_{P1},..n_{PM})
+\left[\sum_{\beta\neq\gamma}^{M}\wp_{N}(n_{P\beta}-n_{P\gamma})-{\cal E}_{M}
\right]\varphi^{(p)}_{M}(n_{P1},..n_{PM})\right\}=0,
\eqno(18)$$
where
$${\cal S}_{\beta}(n_{P1},..n_{PM})=\sum_{s\neq n_{P1},..n_{PM}}^{N}
\wp_{N}(n_{P\beta}-s)\hat Q_{\beta}^{(s)}\varphi_{M}^{(p)}
(n_{P1},..n_{PM}).
\eqno(19)$$
The operator $\hat Q_{\beta}^{(s)}$ in (19) replaces $\beta$th  argument
of the function of $M$ variables to $s$.

To calculate the sum (19), let us introduce, following the consideration
of the hyperbolic exchange in [17], the function of one complex variable
$x$,
$$W_{P}^{(\beta)}(x)=\sum_{s=1}^{M}\wp_{N}(n_{P\beta}-s-x)\hat
Q_{\beta}^{(s+x)}
\varphi_{M}^{(p)}(n_{P1},..n_{PM}).
\eqno(20)$$
As a consequence of (11), (13-14) it obeys the relations
$$W_{P}^{(\beta)}(x+1)=W_{P}^{(\beta)}(x),\qquad
W_{P}^{(\beta)}(x+\omega)=\exp(q_{\beta}(p))W_{P}^{(\beta)}
(x).
\eqno(21)$$
The only singularity of $W_{P}^{(\beta)}$ on the torus $T_{1}={\bf C}/{\bf Z}+
{\bf Z}\omega$ is located at the point $x=0$. It arises from the terms in (20)
with $s=n_{P1},..n_{PM}$. The Laurent decomposition of (20) near $x=0$
has the form
$$W_{P}^{(\beta)}(x)=w_{-2}x^{-2}+w_{-1}x^{-1}+w_{0}+O(x).
\eqno(22)$$
The explicit expressions for $w_{-i}$ can be found from (20),
$$w_{-2}=\varphi_{M}^{(p)}(n_{P1},..n_{PM})
\eqno(23a)$$
$$w_{-1}={\partial\over{\partial n_{P\beta}}}\varphi_{M}^{(p)}(n_{P1},..n_{PM})
$$
$$
+(-1)^P G(n_{1},..n_{M})\sum_{\lambda\neq\beta}T_{\beta\lambda}
(n_{P1},..n_{PM})
\hat Q_{\beta}^{(n_{P\lambda})}
\exp\left(-i\sum_{\nu=1}^M p_{\nu}n_{P\nu}\right)F^{(p)}(n_{P1},..n_{PM})
\eqno(23b)$$
$$w_{0}={\cal S}_{\beta}(n_{P1},..n_{PM})+{1\over2}{{\partial^2}\over
{\partial n_{P\beta}}}\varphi_{M}^{(p)}(n_{P1},..n_{PM})
+(-1)^{P}G(n_{1},..n_{M})$$
$$\times\sum_{\lambda\neq\beta}T_{\beta\lambda}
(n_{P1},..n_{PM})\left[U_{\beta\lambda}(n_{P1},..n_{PM})
\hat Q_{\beta}^{(n_{P\lambda})}
+\wp_{N}(n_{P\beta}-n_{P\lambda})\partial \hat Q_{\beta}^{(n_{P\lambda})}
\right]
\eqno(23c)$$
$$\times\exp\left(-i\sum_{\nu=1}^M p_{\nu}n_{P\nu}\right)
F^{(p)}(n_{P1},..n_{PM}),$$
where
$$T_{\beta\lambda}(n_{P1},..n_{PM})=\sigma_{N}(n_{P\lambda}-n_{P\beta})
\prod_{\rho\neq\beta,\lambda}^{M}
{{\sigma_{N}(n_{P\rho}-n_{P\beta})}\over
{\sigma_{N}(n_{P\rho}-n_{P\lambda})}},$$
$$
U_{\beta\lambda}(n_{P1},..n_{PM})=\wp_{N}'(n_{P\lambda}-n_{P\beta})-
\wp_{N}(n_{P\beta}-n_{P\lambda})
\sum_{\rho\neq\beta,\lambda}\zeta_{N}(n_{P\rho}-n_{P\lambda}),$$
$(-1)^{P}$ means the parity of the permutation $P$ and the action of
the operator
 $\partial\hat Q_{\beta}^{(n_{P\lambda})}$
on the function $Y$ of $M$ variables is defined as
$$\partial Q_{\beta}^{(n_{P\lambda})}Y(z_{1},..z_{M})=
{\partial\over{\partial z_{\beta}}}Y(z_{1},..z_{M})\vert_{z_{\beta}=
n_{P\lambda}}.
\eqno(24)$$
The next step consists in writing the explicit expression
for the function $W_{P}^{(\beta)}(x)$ obeying the relations
(21) and (22) [17],
$$W_{P}^{(\beta)}(x)=\exp(a_{\beta}x)
{{\sigma_{1}(r_{\beta}+x)}\over
{\sigma_{1}(r_{\beta}-x)}}\{w_{-2}(\wp_{1}(x)-\wp_{1}(r_{\beta})+
(w_{-2}(a_{\beta}+2\zeta_{1}(r_{\beta}))-w_{-1})
$$
$$
\times[\zeta_{1}(x-r_{\beta})-\zeta_{1}(x)+\zeta_{1}(r_{\beta})-
\zeta_{1}(2r_{\beta})]\}.
\eqno(25)$$
The Weierstrass functions $\wp_{1},\zeta_{1}$ and $\sigma_{1}$ in (25)
are defined on the torus $T_{1}$ and the parameters $a_{\beta},r_{\beta}$
are chosen as to satisfy the conditions (21),
$$a_{\beta}=(\pi i)^{-1}q_{\beta}(p)\zeta_{1}(1/2)
\qquad r_{\beta}=-(4\pi i)^{-1}q_{\beta}(p).$$
By expanding (25) in powers of $x$ one can find $w_{0}$ in terms of
 $w_{-2},w_{-1},
q_{\beta}$ and obtain the explicit expression for
${\cal S}_{\beta}(n_{P1},..n_{PM})$ with the use of (23a-c). It turns out that
the equation (18)
 can be recast in the form
$$
\sum_{P\in\pi_{M}}\left[-{1\over2}\sum_{\beta=1}^{M}
\left({\partial\over{\partial n_{P\beta}}}-f_{\beta}(p)\right)^2+
\sum_{\beta\neq\gamma}^{M}\wp_{N}(n_{P\beta}-n_{P\gamma})-{\cal E}_{M}+
\sum_{\beta=1}^{M}\varepsilon_{\beta}(p)\right]\varphi^{(p)}(n_{P1},..
n_{PM})$$
$$={1\over2}G(n_{1},..n_{M})\sum_{P\in\pi_{M}}(-1)^{P}
\sum_{\beta\neq\lambda}\left[
Z_{\beta\lambda}(n_{P1},..n_{PM})+Z_{\lambda\beta}(n_{P1},..n_{PM})
\right],
\eqno(26)
$$
where
$$f_{\beta}(p)=(\pi i)^{-1}q_{\beta}(p)\zeta_{1}(1/2)
-\zeta_{1}((2\pi i)^{-1}q_{\beta}(p)),
\eqno(27)
$$
$$\varepsilon_{\beta}(p)={1\over2}\wp_{1}((2\pi i )^{-1}q_{\beta}(p))
\eqno(28)$$
and $Z_{\beta\lambda}(n_{P1},..n_{PM})$ is defined by the relation
$$Z_{\beta\lambda}(n_{P1},..n_{PM})=T_{\beta\lambda}(n_{P1},..n_{PM})
\left[U_{\beta\lambda}(n_{P1},..n_{PM})
\hat Q_{\beta}^{(n_{P\lambda})}+\wp_{N}(n_{P\lambda}-n_{P\beta})\right.
$$
$$\left.\times
(\partial\hat Q_{\beta}^{(n_{P\lambda})}-f_{\beta}(p)
        \hat Q_{\beta}^{(n_{P\lambda})})\right]
\exp\left(-i\sum_{\nu=1}^{M}p_{\nu}n_{P\nu}\right)
F^{(p)}(n_{P1},..n_{PM}).
\eqno(29)$$
Turning to the definition (11) of $\varphi^{(p)}$
one observes that each term of the left-hand side of (26) has the same
structure
as the left-hand side of the many-particle Schr\"odinger equation (12) and
vanishes if ${\cal E}_{M}$ and $f_{\beta}(p)$ are chosen as
$$f_{\beta}(p)=-ip_{\beta},\qquad \beta=1,..M,
\eqno(30)$$
$${\cal E}_{M}={\sf E}_{M}(p)+\sum_{\nu=1}^{M}\varepsilon_{\beta}(p).
\eqno(31)$$
Now let us prove that that the right-hand side of (26) also vanishes.
The crucial observation is that the sum over permutations in it can be recast
in the form
$$\sum_{P\in\pi_{M}}(-1)^{P}\sum_{\beta\neq\lambda}[Z_{
\beta\lambda}(n_{P1},..n_{PM})-Z_{\lambda\beta}(n_{PR1},..n_{PRM})],$$
where $R$ is the transposition $(\beta\leftrightarrow\lambda)$
which leaves other numbers from 1 to $M$ unchanged.
The term in square brackets is simplified drastically with the use of the
identities
$$T_{\lambda\beta}(n_{PR1},..n_{PRM})=T_{\beta\lambda}(n_{P1},..n_{PM}),\quad
  U_{\lambda\beta}(n_{PR1},..n_{PRM})=U_{\beta\lambda}(n_{P1},..n_{PM})$$
$$
\hat Q_{\lambda}^{(n_{P\beta})}F(n_{PR1},..n_{PRM})=
\hat Q_{\beta}^{(n_{P\lambda})}F(n_{P1},..n_{PM}).
$$
Taking into account the relations (29-30), one finds
$$Z_{\beta\lambda}(n_{P1},..n_{PM})-Z_{\lambda\beta}(n_{PR1},..n_{PRM})=
T_{\beta\lambda}(n_{P1},..n_{PM})\wp_{N}(n_{P\lambda}-n_{P\beta})$$
$$\times\exp\left[-i\left((p_{\beta}+p_{\lambda})n_{P\lambda}+
\sum_{\rho\neq\beta,\lambda}^{M}p_{\rho}n_{P\rho}
\right)\right]\left(
{\partial\over{\partial n_{P\beta}}}-
{\partial\over{\partial n_{P\lambda}}}\right)F^{(p)}
(n_{P1},..n_{PM})\vert_{n_{P\beta}=n_{P\lambda}}.
\eqno(32)$$
The last factor in (32) vanishes due to the condition (17)
imposed by the regularity of the left-hand side of the Schr\"odinger equation
(16).

It remains to show that the states of the spin lattice given by (7) with
the functions $\psi_{M}$ of the form (10-11) are highest-weight states with
$S=S_{3}.$ This statement is equivalent to the relation ${\bf S}_{+}\vert
\psi^{(M)}>=0$, which can be rewritten as
$$\sum_{\beta=1}^{M}\sum_{P\in\pi_{M}^{(\beta)}}\sum_{s\neq n_{1},..n_{M-1}}
\hat Q_{\beta}^{(s)}\varphi_{M}^{(p)}(n_{P1},..n_{PM})=0,
\eqno(33)
$$
where $\{\pi_{M}^{(\beta)}\}$ are the subsets of $\pi_{M}$: $P\in\pi_{M}^{
(\beta)}\leftrightarrow P\beta=M$. The sums over $s$ in (33) can be reduced and
presented
in the closed form by using the
technique described above.  It turns out that the left-hand side
of (33) contains the factors similar to the last factor in (32) and vanishes
 due to the condition (17).

The descendant states with $S_{3}< S$ are obtained by acting with ${\bf S}_{-}$
on the basic states $\vert\psi^{(M)}>$ (7). Thus the present consideration
allows, in principle, to reproduce all the eigenvectors of ${\cal H}^{(s)}$
for the exchange (5) as it has been done by Bethe [1] for nearest-neighbor
spin coupling. The equations (30) for the pseudomomenta $\{p\}$ constitute
the analog of the usual Bethe ansatz. The spectrum is given by the relations
(9) and (31).

In conclusion, it is demonstrated that the procedure of the exact
diagonalization of the lattice Hamiltonian with the non-nearest-neighbor
elliptic exchange can be reduced in each sector of the Hilbert space with
given magnetization to the construction of the special double quasiperiodic
eigenfunctions of the many-particle Calogero-Moser problem on a
continuous line. The equations of the Bethe-ansatz form appear very
naturally as a set of restrictions to the particle pseudomomenta. The proof
of this correspodence between lattice and continuum integrable models is
based only on analytic properties of the eigenfunctions. One can expect
that the set of spin lattice states constructed by this way is complete.
This is supported by exact analytic proof in the two-magnon case.

The analysis of explicit form of the equations (30) available for
$M=2,3$ shows that the spectrum of the lattice Hamiltonian with the
exchange (5) is {\it not}
additive being given in terms of pseudomomenta $\{p\}$ or phases which
parametrize the sets $\{p,q\}$ [10,19]. The problem of finding appropriate
set of parameters which gives the "separation" of the spectrum remains
open. It would be also of interest to consider various limits $(N\to
\infty, \kappa\to0,\infty)$ so as to recover the results of the papers
[1,3,17] and prove the validity of the approximate methods of asymptotic
Bethe ansatz after finding explicit form of the functions $q_{\beta}
(p)$ and ${\sf E}_{M}(p)$.

I would like to thank Prof. M. Takahashi for his interest to this work
and useful discussion. The support by the Ministry of Education, Science and
Culture of Japan is gratefully acknowledged.
\newpage
{\bf References}

\begin{enumerate}
\item
 H. Bethe. Z. Phys. 71,205 (1932)
\item
F.D.M. Haldane. Phys. Rev. Lett. 60,635 (1988); B.S. Shastry,
{\it ibid} 60,639 (1988)
\item
F.D.M. Haldane. Phys. Rev. Lett. 66,1529 (1991)
\item
L.D. Faddeev, in {\it Integrable models of 1+1 dimensional
quantum field theory}, Elsevier, Amsterdam, 1984
\item
D. Bernard, M. Gaudin, F.D.M. Haldane and V. Pasquier.
J. Phys.A: Math. Gen. 26, 5219 (1993)
\item
M. Takahashi. Progr. Theor. Phys. 46,401 (1971)
\item
E.H. Lieb and W. Liniger. Phys. Rev. 130,1605 (1963)
\item
M. Takahashi. Progr. Theor. Phys. 91,1 (1994)
\item
N. Kawakami. Progr. Theor. Phys. 91,189 (1994)
\item
V.I. Inozemtsev. J. Stat. Phys. 59,1143 (1990)
\item
F. Calogero. Lett. Nuovo Cim. 13,411 (1975)
\item
J. Moser. Adv. Math. 16,1 (1975)
\item
M.A. Olshanetsky and A.M. Perelomov. Phys. Rep.94,313 (1983)
\item
P.I. Etingof. {\it Quantum integrable systems and representations of
Lie algebras,} hep-th 9311132
\item
P.I. Etingof and A.A. Kirillov Jr. Duke Math. J. 74,585 (1994)
\item
V.I. Inozemtsev. ISSP preprint No. 2928 (1995), to appear in Lett. Math. Phys.
\item
V.I. Inozemtsev. Commun. Math. Phys. 148, 359 (1992)
\item
J. Dittrich and V.I. Inozemtsev. J. Phys.A: Math. Gen. 26, L753
(1993)
\item
V.I. Inozemtsev. ISSP preprint No. 2947 (1995)
\end{enumerate}

\end{document}